%% file: camera_ready.tex
\documentclass{Interspeech2024}

\usepackage{tikz}
\usetikzlibrary{shapes.arrows, arrows.meta,fadings,arrows,matrix,spy, calc}
\usepackage{booktabs}
\usepackage{multirow}
\usepackage{cite}

\makeatletter
\def\bstctlcite{\@ifnextchar[{\@bstctlcite}{\@bstctlcite[@auxout]}}
\def\@bstctlcite[#1]#2{\@bsphack
  \@for\@citeb:=#2\do{%
    \edef\@citeb{\expandafter\@firstofone\@citeb}%
    \if@filesw\immediate\write\csname #1\endcsname{\string\citation{\@citeb}}\fi}%
  \@esphack}
\makeatother

\def\thineq{\hspace{-.1em}=\hspace{-.1em}}

\interspeechcameraready

\title{%
Speech dereverberation constrained on room impulse response characteristics%
}

\name[affiliation={1}]{Louis}{Bahrman}
\name[affiliation={1}]{Mathieu}{Fontaine}
\name[affiliation={2}]{Jonathan}{Le Roux}
\name[affiliation={1}]{Gaël}{Richard}

\address{
  $^1$LTCI, Telecom Paris, Institut polytechnique de Paris, France\\
  $^2$Mitsubishi Electric Research Laboratories (MERL), USA}
\email{$^1$firstname.lastname@telecom-paris.fr, $^2$leroux@merl.com}


\newcommand{\louis}{\textcolor{black}}
 
\newcommand{\jl}{\textcolor{black}}

\input{math_notation}
\begin{document}
\bstctlcite{IEEEexample:BSTcontrol} %

\maketitle

\begin{abstract}

Single-channel speech dereverberation aims at extracting a dry speech signal from a recording %
affected by the acoustic reflections in a room. 
However, most current deep learning-based approaches for speech dereverberation are not interpretable for room acoustics,
and can be considered as black-box systems in that regard. 
In this work, we address this problem by regularizing the training loss using a novel physical coherence loss which encourages the room impulse response (RIR) induced by the dereverberated output of the model
to match the acoustic properties of the room in which the signal was recorded.
Our investigation demonstrates the preservation of the original dereverberated signal alongside the provision of a more physically coherent RIR.

\end{abstract}

\section{Introduction}

An acoustic signal captured in a closed room comprises several correlated components: a more so-called direct-path signal and a combination of early reflections plus late reverberation collectively coined as reverberation signal. 
The reverberation phenomenon may not be desirable in speech recording as it lowers its perceptual intelligibility~\cite{houtgast1985review}. 
This justifies the need to transform the reverberant signal to mitigate its effects in speech-related tasks such as speech enhancement or automatic speech recognition~\cite{yoshioka_making_2012}.
The process of speech dereverberation consists in removing the early reflections and late reverberation from a reverberant signal, thereby approximating the dry signal. 
This presents yet an ill-posed problem since it depends on deconvolution where the impulse response is unknown. 
In theory, the convolutive model used for dereverberation should represent the Room Impulse Response (RIR), which uniquely characterizes reverberation.
As RIR is not minimum-phase~\cite{neely_invertibility_1979} or lacks robustness to spatial variations~\cite{mourjopoulos_variation_1985}, 
a wide range of models mitigate deconvolution errors using regularization of the known RIR~\cite{cahill_novel_2008, kodrasi_frequency-domain_2014}, 
deep generalization to a spatial neighbourhood~\cite{xu_personalized_2023}, 
or by a posterior sampling of a diffusion process informed by the RIR~\cite{lemercier_diffusion_2023}.

A first approach is to directly model either the dry signal, the reverberant signal, or both for dereverberation purposes.
Regarding the modelling of reverberation, it has been represented as a convolutive distortion, and approaches have been developed to concurrently represent the convolutive model and the dry signal~\cite{vincent_audio_2018}.
One of the most notable methods is the Weighted Prediction Error (WPE)~\cite{nakatani_speech_2010}. 
This method has widely benefited from further refinement, including hybrid approaches combining WPE with deep learning~\cite{kinoshita_neural_2017, saito_unsupervised_2023}.
While WPE estimates the time-frequency (T-F) filter used to synthesize a dry signal from a reverberant one, Forward Convolutive Prediction (FCP)~\cite{wang_convolutive_2021_waspaa} aims at estimating the filter mapping a dry signal estimated by a neural network to a reverberant mixture. 
It has been applied to tasks such as dereverberation in a monaural setting~\cite{wang_convolutive_2021}, unsupervised multichannel dereverberation~\cite{wang_usdnet_2024} and source separation~\cite{wang_unssor_2023, aralikatti_reverberation_2023}.

\begin{figure}[t]

   \includegraphics[width=\linewidth]{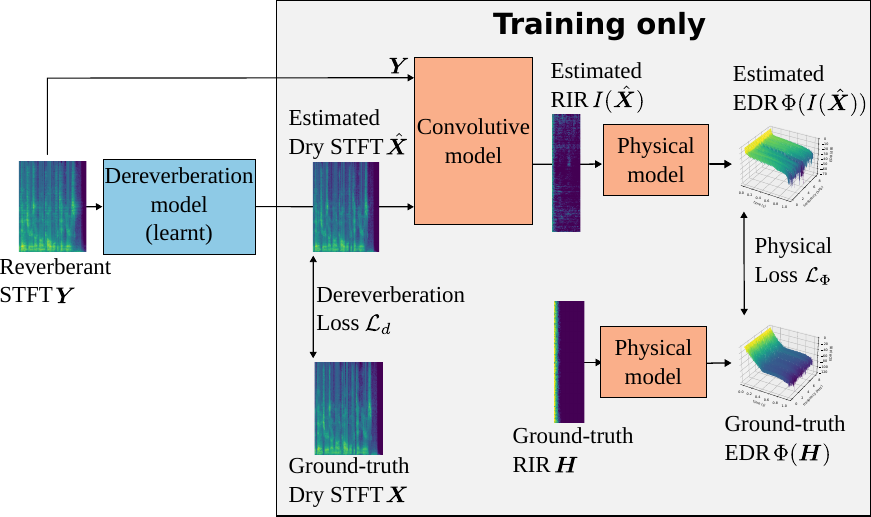}
   \vspace{-4mm}
        \caption{\louis{Overview of the proposed method.}
        }
    \label{fig:overview}
    \vspace{-2mm}
\end{figure}

The FCP is moreover closely related to the Convolutive Transfer Function (CTF) approximation, which considers reverberation as a subband filtering process. 
An observation model based on CTF has been used in conjunction with nonnegative matrix factorization (NMF)~\cite{baby_supervised_2016} and a diffusion model in~\cite{wang_rvae-em_2023}.
However, neither the backward filter estimated by WPE nor the FCP or CTF have been constrained to be realistic with respect to room acoustics. 
Most
state-of-the-art single-channel \jl{DNN-based} dereverberation algorithms such as TF-GridNet~\cite{wang_tf-gridnet_2023} or UNet-based architectures~\cite{ernst_speech_2018} have shown good performance in various scenarios\jl{, yet are} purely data-driven designed.
The dereverberation task can 
\louis{on the other hand}
leverage not only the RIR itself but also the physical properties that constrain it leading to a physics-driven dereverberation paradigm.
This has been made possible by the recent advances in blind room acoustic parameters estimation\louis{~\cite{prego_2015_blind}}.
This approach has first been used to leverage the reverberation time $\rtsixty$ in classical models~\cite{lebart_new_2001}, and refined using DNNs in~\cite{wu_reverberation-time-aware_2017,li_composite_2023} for instance.
At inference, they require a preliminary estimation of the $\rtsixty$ 
but do not constrain the model output to match this property.
\louis{Similarly, }
FullSubnet~\cite{hao_fullsubnet_2021} has been used to target a signal with a shortened $\rtsixty$\louis{~\cite{zhou_speech_2022}}.
\louis{While this physically realistic approach simplifies the learning target of the DNN, the predicted signal is not necessarily dry.}

This paper aims to bridge the gap between convolutive models and room acoustic properties estimation to constrain a deep dereverberation model.
More precisely, for this preliminary study, we choose FullSubNet as a 
\louis{weakly}
physics-driven dereverberation algorithm and design physic\louis{al} losses inherited from a CTF model.
Our contribution is two-fold:
we show that 1) DNNs designed to only dereverberate speech are also able to implicitly model reverberation without increasing the number of parameters and 2) explicitly synthesize an RIR from a dereverberation model.
More precisely, our proposed speech dereverberation constrained on RIR 
procedure demonstrates, through obtained objective scores, that we can maintain the overall quality of the original FullSubNet output while exhibiting a more physically consistent RIR. 
For reproducibility purposes and to help future research, we publicly distribute our code and pretrained models\footnote{\louis{\url{https://louis-bahrman.github.io/SD-cRIRc/}}}.

\section{Reverberation in the T-F domain}
\label{sec:reverb}

{\bf Time-domain formulation:} Assuming fixed source and microphone positions and no additive noise, a monaural reverberant (or wet) signal $y$ can be represented as a convolution between a dry signal $x$ and the room impulse response (RIR) $h$ between the source and the microphone:
\begin{equation}
    y_n = (h * x)_n , \label{eqref:reverb_conv} 
\end{equation}
where $n$ denotes the time index
and $*$ the convolution operator. 

\noindent {\bf STFT filtering and Convolutive transfer function:}
The time-invariant linear system of Eq.~\eqref{eqref:reverb_conv} can be formulated in the short-time Fourier transform (STFT) domain as interband and interframe convolution~\cite{avargel_system_2007}:
\begin{equation}
Y_{f,t}=\sum_{f^{\prime}=0}^{F-1}\sum_{t^{\prime}=-\infty}^{\infty} \mathcal{H}_{f,f^{\prime},t^{\prime}} X_{f^{\prime},t-t^{\prime}}, \label{eq:full_convolution}
\end{equation}
where $Y_{f,t}$ is the STFT coefficient of the reverberant signal at frequency $f\thineq 0,\dots,F-1$ and time $t\thineq 0,\dots,T_y-1$, 
$\mathcal{H} \in \C^{F \times F \times T_h}$ is a tridimensional representation of the RIR 
and ${X} \in \C^{F\times T_x}$ is the $\STFT$ of the dry signal.
As shown in~\cite{avargel_system_2007}, $\mathcal{H}$ can be obtained in closed form from the RIR $h$. 
Several approximations can be made from this model.
Among them, the {subband filtering} operation, also named {convolutive transfer function} (CTF), considers the case where $\mathcal{H}_{f,f',t'}$ is nonzero only if $f=f'$~\cite{vincent_audio_2018}. 
Crossband modelling, investigated in~\cite{avargel_system_2007}, considers an interband convolution kernel $\mathcal{C}_f$ of size 
$(2 F' +1) T_h $
for each frequency band $f$. 
The crossband filter can be estimated from the STFT coefficients of the dry signal (or an estimate of it) and the reverberant signal via a frequency-dependent least-squares optimization problem:
\begin{equation}
    {{\mathcal{C}}}_{f}( %
    \bm{X}) = \argmin_{{\bm{C}}_{f}} \norm{{{\mtx{\bar{X}}}}_f {\bm{C}}_{f} - \bm{Y}_{f} }_2^2, \\ \label{eq:CTF}
\end{equation}
where
\begin{align}
    {\bm{C}}_{f} &\triangleq \left[ \vct{C}^{\mathsf{T}}_{f,f^\prime} \right]_{f^\prime=f-F'}^{f+F'} \in \mathbb{C}^{(2F'+1)T_h}, \\
{\bm{\bar{X}}}_{f} &\triangleq \left[ \mtx{X}^{(\mathrm{T})}_{f^\prime}\right]_{f^\prime=f-F^{\prime}}^{f+F' } \in \mathbb{C}^{T_y \times (2F'+1) T_h}.
\end{align}
${{\mathcal{C}}}_{f}(%
\bm{X})$ is the concatenation of the crossband filters \louis{$\vct{C}^{\mathsf{T}}_{f,f^\prime} $} mapping the frequencies $f' = f-F', \dots, f+F'$ of $\bm{X}$ to the frequency $f$ of the reverberant STFT $\bm{Y}$.
${\bm{\bar{X}}}_{f}$ is the column-wise concatenation of the Toeplitz matrices $\mtx{X}_{f'}^{(\mathrm{T})}$ of size $T_y \times T_h$ constructed from frequency bands $f' \thineq f-F',\dots, f+F^\prime$ of the dry STFT coefficients.

\noindent {\bf Room parameter estimation:}
Given an STFT representation $\bm{H}$ of an impulse response $h$, the energy decay relief (EDR)~\cite{jot_analysissynthesis_1992} is defined for each time-frequency bin $(f,t)$ as:
\begin{equation}
       \EDR(\bm{H})_{f,t} \triangleq \sum_{t'=t}^{+\infty} \lvert H_{f, t^\prime} \rvert^2.
\end{equation}
The EDR can be interpreted as a subband energy decay curve (EDC), representing a frequency-dependent energy decay.
It has been used as a loss for RIR  estimation~\cite{ratnarajah_towards_2023}.

\section{Proposed Method}
\subsection{Overview}

We propose to introduce a new loss term that imposes physical constraints on the RIR characteristics measured via the CTF approximation when training a dereverberation deep neural network (DNN). 
The general procedure to define our physical loss term is as follows. 
From the dereverberated output $\hat{\bm{X}}$ obtained by the DNN from a reverberant signal $\bm{Y}$, a \louis{convolutive model}
computes the CTF ${{\mathcal{C}}}(%
\hat{\bm{X}})$
mapping the output of the DNN to its reverberant input, following Eq.~\eqref{eq:CTF}, and from it an estimate $I(\hat{\bm{X}})$ of the STFT of the corresponding RIR.
\louis{A physical model}
is then used to compute an estimated physical property $\Phi(I(\hat{\bm{X}}))$ from the estimated CTF, and similarly a target physical property $\Phi(I(\bm{X}))$ from the oracle CTF obtained with the ground-truth anechoic signal. Their distance is finally used to define our physical loss function $\L_\phi$.

This new loss term $\L_\phi$ can be combined with a classical dereverberation loss $\L_d$ (e.g., assessing the reconstruction quality of the dry or direct-path signal) to train the DNN.
A diagram of the training procedure is shown in Fig.~\ref{fig:overview}.

Because the
\louis{convolutive and physical}
models are not parametric, they do not need to be trained. 
At inference, for the dereverberation task,  
\louis{these}
blocks are discarded, and only the DNN is used. Hence, the number of parameters, as well as the computational complexity and memory footprint are the same as for the original model. 

\subsection{Corrected Convolutive Model}
The number of crossbands is limited by the dimension of the least-squares system to solve at Eq.~\eqref{eq:CTF}. For the system to have a unique solution, it is required that ${\bm{\bar{X}}}_{f}$ is full-rank, hence the relation $(2F'+1) T_h< T_y$ must hold.
Taking into account the length of the dry signals and RIRs in our training data, as well as the computational load, we limit ourselves to considering the subband ($F'=0$) and 3-band ($F'=1$) cases for the CTF. 
We solve Eq.~\eqref{eq:CTF} using QR decomposition.

It can be proven that the STFT $\bm{H}$ of the impulse response $h$ can be computed from the convolutive interframe and interband filter $\mathcal{H}_{f,f',t}$ (if it were known):
\begin{equation}
{H}_{f,t} = %
\sum_{f'=0}^{F-1} (-1)^{f'} \mathcal{H}_{f,f',t}
\end{equation}
where the multiplication by $(-1)^{f'}$ stems from the centering of the first STFT window.
We can use this relationship to obtain an estimate of the STFT of the RIR from the CTF computed by either the clean speech $\bm{X}$ or its estimate $\hat{\bm{X}}$:
\begin{equation}
    I(\bm{X})_{f,t}=\sum_{f'=f-F'}^{f+F'} (-1)^{f'}\mathcal{C}_{f,f',t}(\bm{X}),
\end{equation}
and similarly for $\hat{\bm{X}}$.
Because our model only considers a few crossbands, this estimate will 
not yield the exact STFT $H_{f,t}$ of the RIR, but an approximation, even if it is computed on the CTF ${\mathcal{C}}(%
\bm{X})$ obtained from the clean speech $\bm{X}$. We define the modeling error at each T-F bin as ${\mathcal{E}}_{f,t} = I( %
\bm{X})_{f,t} - H_{f,t}$.

To make physical properties less dependent on this approximation, we attempt to compensate for the error via a spectral-subtraction-based correction. 
The spectral subtraction yields $I(\bm{X})^c_{f,t}$, an estimator of the RIR spectrum. 
The same error correction can be applied to the estimate $
I(\hat{\bm{X}})$ of the RIR obtained from the estimate $\hat{\bm{X}}$ of the dry speech:
\begin{align}
    I(\bm{X})_{f,t}^c &= \left( \lvert I( \bm{X} )_{f,t}\rvert^2 - \abs{{\mathcal{E}}_{f,t}}^2 \right)^{1/2} e^{j\angle I(\bm{X})_{f,t}},\\
        I(\hat{\bm{X}})_{f,t}^{c} &= \left( \lvert I( \hat{\bm{X}} )_{f,t}\rvert^2 - \abs{{\mathcal{E}}_{f,t}}^2 \right)^{1/2} e^{j\angle I(\hat{\bm{X}})_{f,t}} .\label{eq:spectral_subtraction}
\end{align}
Note that adjusting both target and estimated convolutive transfer functions by the same quantity will alter the nonlinear behaviour of the 
\louis{physical model}
employed.

If the spectrogram of the RIR that has been used for data generation is not available, one can still compare the physical properties estimated from $I(\hat{\bm{X}})$ and $I(\bm{X})$ directly without applying the correction.

\subsection{Physical coherence loss}

As an example of physical characteristic of interest to be used as a constraint on the RIR, we consider the dB-scaled EDR~\cite{jot_analysissynthesis_1992}.
Given an STFT of an RIR or an approximation of it, $\bm{R}$, the dB-scaled EDR is obtained as:
\begin{align}
        \Phi_{f,t}(\bm{R}) \triangleq \EDR^s (\bm{R})_{f,t}= 10\log_{10} \frac{\EDR(\bm{R})_{f,t}}{\EDR(\bm{R})_{f,0}}.
\end{align}
The physical coherence loss $\L_\Phi$ can then be defined as a point-wise mean-squared error between the dB-scaled EDRs obtained from an estimate $\hat{\bm{R}}$ and a target $\bm{R}$. 
Since the tail of the EDR is very sensitive to CTF approximation errors and has high values on the log scale, 
both target and estimated EDRs are masked to exclude time-frequency bins where the target EDR is lower than $-20$ dB:
\begin{multline}
\L_\Phi(\hat{\bm{R}},\bm{R})=\sum_{f,t} \big|{\Phi}_{f,t}(\hat{\bm{R}}) - \Phi_{f,t}(\bm{R})\big|^2 \mathds{1}_{\{ \Phi_{f,t}(\bm{R})> -20\}} .
\end{multline}
We consider several variants for the selection of $\hat{\bm{R}}$ and $\bm{R}$, such as $I(\hat{\bm{X}})^c$ and $I(\bm{X})^c$, as described in Section~\ref{sec:variants}.

\subsection{Multi-objective training}
To balance both physical coherence and reconstruction losses in a multi-task training setting, we use GradNorm~\cite{chen_gradnorm_2018}.
GradNorm ensures that the gradients of both $\L_\Phi$ and $\L_d$ losses have equal norms across all weights.
In our setting, $\L_\Phi$ is highly nonconvex with respect to the network parameters, so we prioritize the reconstruction loss over the physical coherence loss to stabilize training. After GradNorm has been applied, we further multiply the physical coherence loss by a constant weight $w_\Phi$. Based on preliminary experiments, we set $w_\Phi = 0.1$.

\section{Experiments}
\begin{table*}[t]
\centering
\sisetup{
detect-weight, %
mode=text, %
tight-spacing=true,
round-mode=places,
round-precision=2,
table-format=1.2,
table-number-alignment=center
}
\caption{Dereverberation scores $\pm$ standard deviation (std.) for FullSubNet (FSN) and its constraints versions.}
\vspace{-1mm}
\label{tab:res_dereverb}
\resizebox{\textwidth}{!}{
\setlength{\tabcolsep}{3pt}
\begin{tabular}{l *{4}{S@{\,\( \pm \)\,}SS[round-precision=1,table-format=2.1]@{\,\( \pm \)\,}S[round-precision=1,table-format=1.1]S@{\,\( \pm \)\,}S}}
& \multicolumn{12}{c}{Matched RIRs} & \multicolumn{12}{c}{Mismatched RIRs} \\
\cmidrule(lr){2-13} \cmidrule(lr){14-25}
& \multicolumn{6}{c}{WSJ0} &  \multicolumn{6}{c}{LibriSpeech clean}& \multicolumn{6}{c}{WSJ0} & \multicolumn{6}{c}{LibriSpeech clean} \\
\cmidrule(lr){2-7} \cmidrule(lr){8-13} \cmidrule(lr){14-19} \cmidrule(lr){20-25}
& \multicolumn{2}{c}{STOI} & \multicolumn{2}{c}{SISDR} & \multicolumn{2}{c}{WB-PESQ}& \multicolumn{2}{c}{STOI} & \multicolumn{2}{c}{SISDR}& \multicolumn{2}{c}{WB-PESQ}& \multicolumn{2}{c}{STOI} & \multicolumn{2}{c}{SISDR}& \multicolumn{2}{c}{WB-PESQ}& \multicolumn{2}{c}{STOI} & \multicolumn{2}{c}{SISDR}& \multicolumn{2}{c}{WB-PESQ}\\
\midrule
FSN & 0.927 & 0.066 & \bfseries 5.106 & \bfseries 4.089 & 2.227 & 0.598 & 0.897 & 0.109 & \bfseries 3.111 & \bfseries 4.340 & 2.061 & 0.551 & 0.871 & 0.059 & \bfseries 0.864 & \bfseries 2.577 & 1.599 & 0.206 & 0.837 & 0.100 & \bfseries -0.772 & \bfseries 3.351 & 1.534 & 0.243\\  
+ SB & 0.923 & 0.068 & 4.326 & 4.245 & 2.104 & 0.559 & 0.894 & 0.110 & 2.512 & 4.587 & 1.983 & 0.507 & 0.859 & 0.063 & -0.313 & 2.879 & 1.457 & 0.188 & 0.823 & 0.102 & -1.853 & 3.547 & 1.420 & 0.214 \\  
+ CSB & 0.922 & 0.069 & 4.187 & 4.641 & 2.114 & 0.646 & 0.894 & 0.111 & 2.222 & 5.098 & 1.985 & 0.585& 0.856 & 0.063 & -0.721 & 2.936 & 1.430 & 0.184 & 0.822 & 0.101 & -2.352 & 3.790 & 1.408 & 0.211 \\  
+ SSB & 0.926 & 0.065 & 4.803 & 4.082 & 2.190 & 0.587 & 0.893 & 0.109 & 2.628 & 4.543 & 1.987 & 0.524 & 0.871 & 0.059 & 0.579 & 2.696 & 1.573 & 0.200 & 0.834 & 0.100 & -1.297 & 3.834 & 1.489 & 0.228 \\  
+ 3B & 0.926 & 0.066 & 4.897 & 4.093 & \bfseries 2.237 & \bfseries 0.595 & 0.895 & 0.110 & 2.919 & 4.605 & \bfseries 2.074 & \bfseries 0.571 & 0.870 & 0.060 & 0.667 & 2.636 & \bfseries 1.614 & \bfseries 0.211 & 0.836 & 0.100 & -1.036 & 3.720 & \bfseries 1.543 & \bfseries 0.249\\ 
\midrule
input & 0.862 & 0.090 & -0.153 & 4.795 & 1.764 & 0.671 & 0.849 & 0.121 & -0.967 & 5.492 & 1.892 & 0.757 & 0.751 & 0.069 & -4.498 & 2.947 & 1.195 & 0.106 & 0.736 & 0.099 & -5.221 & 3.749 & 1.242 & 0.160\\  
\end{tabular}
}
\end{table*}

\subsection{Model variants}
\label{sec:variants}
We assess several variants of our method with FullSubNet (FSN)~\cite{hao_fullsubnet_2021}
as the baseline dereverberation model (see Fig.~\ref{fig:overview}).
The ability of FullSubNet to process spectrograms both in the full-band and subband directions is required to estimate a cross-band convolutive model.
It has also been successfully used to solve the physically meaningful task of reverberation-time shortening~\cite{zhou_speech_2022}. 
We select its bidirectional version and keep the original training loss expressed as a mean square error on its complex ratio mask output
\cite{williamson_complex_2016} as the dereverberation loss $\mathcal{L}_d$.

The following variants are considered, representing different ways to compute the convolutive model. 
We define two kinds of approaches depending on whether subband or crossband filters are considered to obtain the estimates of the RIR STFT, and which estimates and targets are compared:

\begin{itemize}
    \item Subband approach (SB):
    $\mathcal{L}_\Phi(I(\hat{\bm{X}}),\bm{H})$, comparing the estimate from $\hat{\bm{X}}$ with ground-truth RIR STFT $\bm{H}$.
    \item Symmetric Subband approach (SSB): $\mathcal{L}_\Phi(I(\hat{\bm{X}}),I({\bm{X}}))$, comparing the estimate from $\hat{\bm{X}}$ with the estimate from $\bm{X}$.
    \item Corrected Subband approach (CSB): $\mathcal{L}_\Phi(I(\hat{\bm{X}})^c,I({\bm{X}})^c)$, comparing the corrected estimate from $\hat{\bm{X}}$ with the corrected estimate from $\bm{X}$.
    \item 3-band approach (3B): $\mathcal{L}_\Phi(I(\hat{\bm{X}}),\bm{H})$, similar to SB but computed using $F'=1$ crossbands.    
\end{itemize}

\subsection{Miscellaneous configurations}
As in the original FullSubNet, 49151 sample excerpts (around 3 s at 16 kHz) reverberant audios are processed in the STFT domain using a 512-sample Hann window with an overlap of 50~\%.
The network is trained for $3\louis{3}0,000$ steps using the Adam optimizer with an initial learning rate (LR) of $10^{-4}$ and a One-cycle-LR with a maximum at $10^{-3}$.

\subsection{Training dataset}
\label{sec:training_dataset}
Similarly to~\cite{wang_tf-gridnet_2023}, we simulated a training dataset by dynamically convolving dry speech signals with simulated RIRs.
The dry speech signals are randomly sampled from the close-talking microphone recordings in the WSJ0 dataset~\cite{wsj0}.
The training set is composed of a total of 61 hours of recordings split into 31,350 audio excerpts.
The simulated RIR dataset consists of 32,000 RIRs simulated using the pyroomacoustics library~\cite{scheibler_pyroomacoustics_2018}
with 2000 rooms whose 
dimensions
 and $\rtsixty$ are uniformly sampled in the respective ranges of $[5,10] \times [5,10] \times [2.5,4]~\text{m}^3$, and $[0.2,1.0]$~s.
In each room, a source is randomly positioned and 16 microphones are sampled such that the source-microphone distance $D$ is uniformly distributed in $[0.75, 2.5]$~m and both source and microphone are at least $50$~cm from the walls.
At training time, we use a dynamic mixing procedure consisting in randomly selecting a dry signal and RIR pair. In order to align the dry signal target and the direct-path, the samples before the direct path are discarded and it is normalised (so that the first impulse is of amplitude 1). This does not change the RIR distribution and 
compensates for the delay induced by the direct-path, both on the STFT $\bm{H}$ and on the oracle EDR $\Phi(\bm{H})$ so that they will start decreasing at the first frame.

We evaluate the proposed method on two different tasks: speech dereverberation and room impulse response characterization.

\subsection{Metrics for evaluation and tasks}
We evaluate the generalization performance of our metrics to both unseen sources and rooms. For dry sources, we consider
the test set of WSJ0~\cite{wsj0}, and Librispeech clean~\cite{panayotov_librispeech_2015}. Two reverberation datasets are considered: one simulated using unseen rooms matching the same physical parameters 
\louis{as the training dataset described in Section~\ref{sec:training_dataset}}
("Matched RIRs"), and the other matching harder conditions ("Mismatched RIRs"): $\rtsixty \in [1.0,1.5]$~s room size range in $[10,15] \times[10,15] \times [4,6]~\text{m}^3$, $D\in [2.5, 4.0]$~m. 
The dereverberation performance between the baseline and the proposed approaches is evaluated using the Short-time-objective Intelligibility STOI, the Scale Invariant Signal-to-noise ratio (SISDR)~\cite{roux_sdr_2019}, and the wide-band Perceptual Evaluation of Speech Quality WB-PESQ.

To demonstrate the acoustic matching capability acquired by the network constrained by RIR characteristics, we compare the energy decay curves (EDCs) predicted at the output of each version of the DNN using 3 convolutive models: 
\begin{itemize}
    \item \jl{EDC-Fourier}: $\L_\Phi \left(\IDFT\left[ \frac{\DFT(y_n)}{\DFT(x_n)} \right], h_n\right)$,\\
    \louis{$\text{where (I)DFT}$ is the (inverse) discrete Fourier transform, }
    and $\Phi_n(r) = \sum_{n'=n}^{+\infty} \abs{r(n')}^2$
   ~\cite{zhou_speech_2022}.
    \item EDR-Subband: Corresponds to the SB loss, which we consider as a metric.
    \item EDR-Crossband: Corresponds to the 3B loss, which we consider as a metric.
\end{itemize} 

\section{Results and Discussion}

\subsection{Dereverberation}

The results for the dereverberation task are presented in Table~\ref{tab:res_dereverb}.
\louis{Our proposed solution, FSN+3B, has a higher WB-PESQ on all datasets and acoustic conditions than the FSN baseline.}
All physically constrained variants exhibit similar performance in terms of STOI 
as the baseline. 
This means that the physical coherence loss and the dereverberation loss can be jointly optimized and that they both converge to equally performing optima in terms of STOI on the space of the DNN weights. 
The poorer results \louis{of our methods compared to the baseline} in terms of \louis{SISDR} can be explained by the DNN \louis{encountering}
difficulty in optimizing the phase of the complex mask mapping $\bm{Y}$ to $\bm{X}$ when it is constrained by a convolutive model.
\louis{Considering this metric}, the model trained on SSB performs similarly to the model trained on 3B.
These losses are the ones that introduce the least constraints on the training and that are the least well-defined
(\louis{S}SB by introducing subband modelling errors, and 3B by being unstable). 
Because these two losses \louis{regularize}
the training in a physically realistic manner, they enable the model to perform better on unseen cases and to generalize to out-of-domain RIRs and source signals. 
Further experiments show that the dereverberation performance remains consistent when high SNR noise is added to the reverberant input of the model at test time.
These results reflect FullSubNet's underlying design assumption that both Full- and Subband modelling \louis{are}
needed for the dereverberation task. 

\subsection{RIR estimation}

\begin{table}[]
\centering
\sisetup{
detect-weight, %
mode=text, %
tight-spacing=true,
round-mode=places,
round-precision=1,
table-format=2.1,
table-number-alignment=center
}
\setlength{\tabcolsep}{4pt}
\caption{RIR estimation scores $\pm$ std. on the WSJ0 test set.}
\vspace{-1mm}
\label{tab:res_rir}
\resizebox{1.0\linewidth}{!}{
    \begin{tabular}{l *{6}{S@{\,\( \pm \)\,}S[round-precision=0,table-format=2]}}
         & \multicolumn{6}{c}{Matched RIRs} & \multicolumn{6}{c}{Mismatched RIRs}\\
         \cmidrule(lr){2-7} \cmidrule(lr){8-13}
         & \multicolumn{2}{c}{EDC} & \multicolumn{4}{c}{EDR}& \multicolumn{2}{c}{EDC} & \multicolumn{4}{c}{EDR}\\
         \cmidrule(lr){2-3} \cmidrule(lr){4-7} \cmidrule(lr){8-9}  \cmidrule(lr){10-13} 
         & \multicolumn{2}{c}{Fourier}& \multicolumn{2}{c}{Subband}& \multicolumn{2}{c}{Crossband}& \multicolumn{2}{c}{Fourier}& \multicolumn{2}{c}{Subband}& \multicolumn{2}{c}{Crossband}\\
         \hline
FSN & 66.201 & 27.553 & 38.979 & 11.908 & 99.616 & 24.255 & 86.382 & 14.933 & 37.815 & 7.225 & 116.669 & 6.455 \\  
+SB & 60.507 & 20.571 & \bfseries 32.743 & \bfseries 6.791 & 100.701 & 22.373 & 66.299 & 16.121 & 27.598 & 5.550 & 114.857 & 6.703 \\  
+CSB & \bfseries 52.551 & \bfseries 23.642 & 34.145 & 12.827 & \bfseries 97.767 & \bfseries 23.702 & \bfseries 63.087 & \bfseries 16.158 & \bfseries 25.568 & \bfseries 4.394 & \bfseries 113.578 & \bfseries 6.803 \\  
+SSB & 76.357 & 23.473 & 39.947 & 9.779 & 102.905 & 23.452 & 86.227 & 13.786 & 40.413 & 7.769 & 117.895 & 6.146 \\  
+3B & 67.065 & 27.472 & 38.708 & 11.446 & 100.034 & 24.242 & 86.775 & 15.059 & 37.471 & 7.002 & 117.161 & 6.196\\
\midrule
dry & 0.000 & 0.001 & 36.687 & 10.227 & 74.982 & 18.914 & 0.000 & 0.001 & 38.361 & 8.469 & 84.381 & 12.146 \\  
    \end{tabular}
    }
\end{table}

Table \ref{tab:res_rir} compares the performance of all proposed approaches %
with respect to the energy decay of several convolutive models. 
The line denoted 
"dry"
shows 
\louis{$\mathcal{L}_\Phi(I(\bm{X}),\bm{H})$}
for each convolutive model and energy decay EDC-Fourier, EDR-Subband, and EDR-Crossband.  It represents the best theoretical performance each convolutive model can offer. 
The results of the 3B metric show a very high error and variance. This can be explained through  Avargel's error analysis of the Crossband filtering~\cite{avargel_system_2007}, where it is shown that for a given SNR on the dry and reverberant signal, there exists only one single tuple $(F', T_x)$ minimizing the mean-squared error. Further analysis shows that the length of the signal considered was too short for the Crossband method to perform well, hence its poor results.
\louis{The results suggest that the RIR estimation task competes with the dereverberation task, as indicated by their differing performance rankings.}
The FSN+CSB variant is performing the best and is capable of modelling the subband model even better than the oracle subband model ${I}(\bm{X})$.
This can be explained by the fact that forcing the model output to respect a \louis{s}ubband model while maintaining its ability to process crossbands in its latent representation is very efficient to predict the STFT, but insufficient to perform dereverberation correctly. This assumption is indeed at the core of FullSubNet's design.
\louis{Accordingly, a general guideline might be to resort to FSN+CSB for the RIR estimation task, and to FSN+3B for the dereverberation task.}

\section{Conclusion}
We have proposed a novel approach for speech dereverberation which constrains the estimated room impulse response to well capture the acoustic properties of the room in which the signal was recorded.
While the overall dereverberation performance remains comparable to the baseline model, having access to a realistic room impulse response characterizing the reverberated environment opens the path to a variety of controllable acoustic transformation applications (acoustic sound matching, realistic room shape modifications,...). 
Future work will be dedicated to the generalization of our approach to other DNN architectures.

\section{Acknowledgements}

This work was funded by the European Union (ERC, HI-Audio, 101052978). Views and opinions expressed are however those of the author(s) only and do not necessarily reflect those of the European Union or the European Research Council. Neither the European Union nor the granting authority can be held responsible for them.
This work was performed using HPC resources from GENCI–IDRIS (Grant 2024-AD011014072R1).
We would like to thank  the reviewers and meta reviewers for their insightful comments.

\bibliographystyle{IEEEtran}
\bibliography{mybib} %

\end{document}

%% file: math_notation.tex
\usepackage{amsmath, amssymb, amsfonts}
\usepackage{dsfont}
\usepackage{bm}

\def\C{\mathbb{C}}

\def\L{\mathcal{L}}

\newcommand{\vct}[1]{\bm{#1}}
\newcommand{\mtx}[1]{\bm{#1}}
\newcommand{\norm}[1]{\left\lVert#1\right\rVert}
\newcommand{\abs}[1]{\left\lvert#1\right\rvert}

\DeclareMathOperator*{\argmin}{arg\,min}

\DeclareMathOperator{\DFT}{DFT}
\DeclareMathOperator{\IDFT}{IDFT}
\DeclareMathOperator{\STFT}{STFT}

\DeclareMathOperator{\EDR}{EDR}

\def\rtsixty{\text{RT}_{60}}

\usepackage{xcolor}
\definecolor{mpc1}{rgb}{0.1216, 0.4667, 0.7059}
\definecolor{mpc2}{rgb}{1.0, 0.498, 0.0549}
\definecolor{mpc3}{rgb}{0.1725, 0.6275, 0.1725}
\definecolor{mpc4}{rgb}{0.8392, 0.1529, 0.1569}
\definecolor{mpc5}{rgb}{0.5804, 0.4039, 0.7412}
\definecolor{mpc6}{rgb}{0.549, 0.3373, 0.2941}
\definecolor{mpc7}{rgb}{0.8902, 0.4667, 0.7608}
\definecolor{mpc8}{rgb}{0.498, 0.498, 0.498}
\definecolor{mpc9}{rgb}{0.7373, 0.7412, 0.1333}
\definecolor{mpc10}{rgb}{0.0902, 0.7451, 0.8118}